\documentclass[a4paper,fleqn,usenatbib]{mnras}

\usepackage[T1]{fontenc}
\usepackage{ae,aecompl}

\usepackage{graphicx}	% Including figure files
\usepackage{amsmath}	% Advanced maths commands
\usepackage{amssymb}	% Extra maths symbols

\usepackage{txfonts}

\title[3$d$ cosmic shear and the iSW effect]{The cross-correlation between 3$d$ cosmic shear and the integrated Sachs-Wolfe effect}

\author[B. Zieser, Ph. M. Merkel]
{Britta Zieser\thanks{E-mail: zieser@stud.uni-heidelberg.de} and Philipp M. Merkel
\\
Universit\"at Heidelberg, Zentrum f\"ur Astronomie, Institut f\"ur Theoretische Astrophysik, Philosophenweg 12, 69120 Heidelberg, Germany}

\date{\today}

\pubyear{2016}

\begin{document}
\label{firstpage}
\pagerange{\pageref{firstpage}--\pageref{lastpage}}
\maketitle

\begin{abstract}
We present the first calculation of the cross-correlation between three-dimensional cosmic shear and the integrated Sachs-Wolfe (iSW) effect. Both signals are combined in a single formalism, which permits the computation of the full covariance matrix. In order to avoid the uncertainties presented by the non-linear evolution of the matter power spectrum and intrinsic alignments of galaxies, our analysis is restricted to large scales, i.e. multipoles below $\ell=1000$. We demonstrate in a Fisher analysis that this reduction compared to other studies of three-dimensional weak lensing extending to smaller scales is compensated by the information that is gained if the additional iSW signal and in particular its cross-correlation with lensing data are considered. Given the observational standards of upcoming weak lensing surveys like \textit{Euclid}, marginal errors on cosmological parameters decrease by ten per cent compared to a cosmic shear experiment if both types of information are combined without a CMB prior. Once the constraining power of CMB data is added, the improvement becomes marginal.
\end{abstract}


%%%%%%%%%%%%%%%%%%%%%%%%%%%%%%%%%%%%%%%%%%%%%%%%%%

%%%%%%%%%%%%%%%%% BODY OF PAPER %%%%%%%%%%%%%%%%%%

\section{Introduction}
Gravitational lensing by the large-scale structure of the Universe gives rise to the cosmic shear field, the statistics of which probe the matter power spectrum and its time evolution as well as spacetime geometry. The distortion of a galaxy image depends on the projected mass along the light path, and therefore varies with the direction of the line-of-sight, i.e. the position of the galaxy on the sky. If galaxy ellipticities are measured across large fields, spatial patterns can be analysed and statistics of the shear field extracted. The sensitivity to cosmological parameters was highlighted by \citet{1996MNRAS.281..369V, 1997A&A...322....1B, 1998ApJ...498...26K, 1999A&A...342...15V, 2002PhRvD..65f3001H}. In addition, the shear imprinted on a galaxy image depends on the distance of the source: the line-of-sight is longer for more remote galaxies, and the different geometric arrangement alters the lensing efficiency of foreground structures. If the shear is treated as a three-dimensional variable, it can constrain the distribution and growth of structures as well as the distance-redshift relation. Fully three-dimensional treatments were introduced by \citet{2003MNRAS.343.1327H} and further developed and applied by \citet{2005PhRvD..72b3516C,2006MNRAS.373..105H,2007MNRAS.376..771K, 2011MNRAS.413.2923K, 2012MNRAS.422.3056A, 2014MNRAS.437.2632G, 2014MNRAS.442.1326K}. To exploit the full spatial dependence, however, adequate redshift measurements are required, which have only recently begun to be available. In absence of such data, only the  statistics of the two-dimensional shear field are accessible \citep{1997ApJ...484..560J, 2002MNRAS.337..875T, 2003MNRAS.340..580T, 2006A&A...452...63M, 2013ApJ...765...74J,2013MNRAS.430.2200K}. Observables are constructed from a projection of the lensing signal along the line-of-sight. In the course of this projection, however, information about the redshift evolution is inevitably lost. As a compromise, source galaxies may be assigned to bins according to their estimated redshifts. In each bin a conventional two-dimensional analysis can be carried out; the range of the projection, however, is much smaller, and the correlations between signals in different redshift bins put constraints on the time evolution. This stacking approach has been termed \lq tomography\rq{} \citep{1999ApJ...522L..21H,2004ApJ...601L...1T,2004A&A...417..873S,2004MNRAS.348..897T,2009JCAP...04..012H,2009A&A...497..677K,2012MNRAS.423.3445S,2013MNRAS.432.2433H}. Several authors have focused in particular on the potential of weak lensing to constrain dark energy \citep{2003PhRvL..91n1302J,2004ApJ...600...17B,2006JCAP...06..025H,2008JCAP...04..013A,2010GReGr..42.2177H}, leading to slighly weaker constraints compared to 3d-methods.

Most recently cross-correlations between the cosmic microwave background (CMB) and cosmic shear have come into focus. Like the light from distant galaxies, the CMB is deflected by the large-scale structure, leading to a remapping of the temperature distribution \citep{2006PhR...429....1L}; in contrast to cosmic shear, only a single source plane (the surface of last scattering) is involved. Cross-correlations between two-dimensional cosmic shear maps and CMB lensing were measured for the first time by \citet{2015PhRvD..91f2001H}. \citet{2014PhRvD..89f3528T} and \citet{2014MNRAS.443L.119H} investigated the influence of intrinsic alignments on this measurement. \citet{2015MNRAS.449.2205K} presented a formalism combining the three-dimensional galaxy ellipticity field and the CMB, considering the temperature, deflection and polarisation fields. They provided error forecasts for cosmological parameters based on the individual data sets and the combination. Expected constraints on the dark energy equation of state were shown to improve by $10$--$15$ per cent when correlations between weak lensing of galaxies and the CMB deflection were considered.

We study cross-correlations between cosmic shear and the integrated Sachs-Wolfe (iSW) effect 
\citep[see the review by][]{2014PTEP.2014fB110N}. Photons crossing a time-evolving potential experience a net frequency shift, reflected in a small change of the CMB temperature. To disentangle the fluctuations from other sources of anisotropy despite their small magnitude, the cross-correlation between the CMB temperature and a suitable tracer is measured \citep{2002PhRvD..65j3510C}. Such a quantity should mark the potential wells, which give rise to the frequency shifts. Detections have been based on the X-ray background \citep{2004Natur.427...45B,2006MNRAS.365..171G}, the galaxy density in various filters \citep{2003ApJ...597L..89F,2005PhRvD..72d3525P,2006MNRAS.365..891V,2007MNRAS.376.1211M,2007MNRAS.377.1085R,2008PhRvD..77l3520G} and the distributions of radio sources \citep{2004Natur.427...45B,2006PhRvD..74d3524P,2008MNRAS.386.2161R} and quasars \citep{2006PhRvD..74f3520G}. Recently the iSW effect was measured from CMB data alone through the cross-correlation of temperature and lensing maps \citep{2015arXiv150201595P}.  The effect vanishes in the linear regime of structure growth during matter domination, so that observations probe the relatively late times at which the dark energy contribution to the density becomes appreciable. This cosmological sensitivity was demonstrated by \citet{2006MNRAS.365..891V,2006PhRvD..74d3524P,2007MNRAS.376.1211M,2008PhRvD..77l3520G,2008PhRvD..78d3519H,2014PhRvD..89b3511G,2015arXiv150201595P,2016MNRAS.456..109M}.

Like cosmic shear, the iSW effect traces potential wells projected along the line-of-sight, albeit with a different radial weight function, and therefore correlations between the two measurements are expected. The iSW effect can in fact be described as a second order lensing phenomenon by relating it to the gravitomagnetic vector potentials, as demonstrated by \citet{2006MNRAS.369..425S} \citep[see also][]{2013MNRAS.431.2433M}. We show how information from both effects can be combined in a single formalism, tying into the three-dimensional treatment of weak lensing. The 3d-method presented here combines iSW-lensing correlations \citep{2008ApJ...676L..93N, 2015arXiv151004770L} in a redshift-resolved way, improving over tomographic methods \citep{2008PhRvD..78d3519H, 2012MNRAS.425.2589J}.

\section{3$d$ formalism}\label{sec:formalism}

\subsection{Shear signal}
The shear tensor $\gamma(\chi,\hat{\mathbf{n}})$ is defined as the second~$\eth$-derivative of the lensing potential $\phi(\chi,\hat{\mathbf{n}})$ \citep{2003MNRAS.343.1327H,2005PhRvD..72b3516C},
\begin{equation}\label{eq:wl_gamma}
  \gamma(\chi,\hat{\mathbf{n}})=\frac{1}{2}\eth\eth\:\phi(\chi,\hat{\mathbf{n}}),
\end{equation}
which in turn is a weighted line of sight-projection of the gravitational potential:
\begin{equation}\label{eq:wl_phi}
  \phi(\chi,\hat{\mathbf{n}}) = 2\int_0^{\chi}\mathrm{d}\chi^{\prime}\,\frac{\chi-\chi^{\prime}}{\chi\chi^{\prime}}\frac{\Phi(\chi,\hat{\mathbf{n}})}{c^2},
\end{equation}
out to the comoving distance $\chi$. Here spatial flatness has been assumed, and the integration is carried out in Born's approximation, i.e. along the unperturbed light path. In order to analyse the shear field in spherical coordinates $(\chi\hat{\mathbf{n}},\chi)$ and decompose it into modes we use the spin-weighted spherical harmonics ${_s}Y_{\ell m}(\hat{\mathbf{n}})$ \citep{NewmanPenrose} as a natural choice due to the spin-2 property of the weak lensing shear,
\begin{equation}\label{eq:spin_transform}
  \gamma_{\ell m}(k)=\sqrt{\frac{2}{\pi}}\int\chi^2\,\mathrm{d}\chi\int\mathrm{d}\Omega\,\gamma(\chi,\hat{\mathbf{n}})j_{\ell}(k\chi){}_2Y_{\ell m}^{\ast}(\hat{\mathbf{n}}).
\end{equation}
For the radial degree of freedom, the spherical Bessel functions~$j_{\ell}(x)$ \citep{1988AmJPh..56..958A} provide a suitable basis, defining the continuous wave number~$k$, whereas the indices $\ell$ and $m$ define the discrete angular wave numbers. Applying an expansion into spherical waves on $\Phi(\chi,\hat{\mathbf{n}})$ yields the expression
\begin{multline}
  \gamma(\chi,\hat{\mathbf{n}})=\sqrt{\frac{2}{\pi}}\frac{1}{c^2}\int_0^{\chi}\mathrm{d}\chi^{\prime}\,\frac{\chi-\chi^{\prime}}{\chi\chi^{\prime}}\\\times
  \int\mathrm{d}k\,k^2\sum_{\ell m}\sqrt{\frac{(\ell+2)!}{(\ell-2)!}}\Phi_{\ell m}(k;\chi^{\prime})j_{\ell}(k\chi^{\prime}){}_2Y_{\ell m}(\hat{\mathbf{n}}),
\end{multline}
where the dependence of the gravitational potential on comoving distance $\chi$ determines the time-evolution of $\Phi_{\ell m}(k)$ due to structure formation. 

The gravitational potential~$\Phi_{\ell m}(k;\chi)$ can be linked to the density contrast~$\delta_{\ell m}(k;\chi)$ using Poisson's equation,
\begin{equation}\label{eq:PoissonsEq}
  \frac{\Phi_{\ell m}(k;\chi)}{c^2} = -\frac{3}{2}\frac{\Omega_{\mathrm{m}}}{(k\chi_\mathrm{H})^2}\frac{\delta_{\ell m}(k;\chi)}{a(\chi)},
\end{equation}
where $\chi_\mathrm{H}\equiv c/H_0$ is the Hubble radius, and a corresponding statistical description in terms of the cold dark matter (CDM) spectrum $P_\delta^0(k)$,
\begin{equation}
  \left\langle\delta_{\ell m}^0(k)\delta^{0*}_{\ell^{\prime}m^{\prime}}(k^{\prime})\right\rangle=\frac{P_{\delta}^0(k)}{k^2}\delta^{\mathrm{D}}(k-k^{\prime})\,\delta^{\mathrm{K}}_{\ell\ell^{\prime}}\delta^{\mathrm{K}}_{m m^{\prime}}.
\end{equation}
Here~$\delta_{\ell m}^0(k)$ denotes the amplitude of a given mode evolved linearly to the present epoch,
\begin{equation}\label{eq:density_decoupled}
  \delta_{\ell m}(k;\chi)=D_+\left[a(\chi)\right]\delta_{\ell m}^0(k),
\end{equation}
 and the raised asterisk marks the complex conjugate. The covariance of the weak lensing shear modes $\gamma_{\ell m}(k)$ can then be traced to the linear power spectrum today, and the full result for the covariance of cosmic shear modes is
\begin{multline}\label{eq:shear_auto}
  \left\langle\bar{\gamma}_{\ell m}(k)\bar{\gamma}_{\ell^{\prime}m^{\prime}}^*(k^{\prime})\right\rangle=\frac{9\Omega_{\mathrm{m}}^2}{16\pi^4\chi_{\mathrm{H}}^4}\frac{(\ell+2)!}{(\ell-2)!}\\\times
  \int\mathrm{d}\tilde{k}\,\frac{P_{\delta}^0(\tilde{k})}{\tilde{k}^2}G_{\ell}(k,\tilde{k})G_{\ell}(k^{\prime},\tilde{k}) \, \delta_{\ell\ell^{\prime}}^{\mathrm{K}}\delta_{mm^{\prime}}^{\mathrm{K}}.
\end{multline}
Here we have introduced the following functions:
\begin{align}
  \label{eq:covariance_G}G_{\ell}(k,k^{\prime})&=\int\mathrm{d}z\,n_z(z)F_{\ell}(z,k)U_{\ell}(z,k^{\prime}),\\
  \label{eq:covariance_F}F_{\ell}(z,k)&=\int\mathrm{d}z_{\mathrm{p}}\,p(z_{\mathrm{p}}|z)j_{\ell}\left[k\chi^0(z_{\mathrm{p}})\right],\\
  \label{eq:covariance_U}U_{\ell}(z,k)&=\int_0^{\chi(z)}\mathrm{d}\chi^{\prime}\,\frac{\chi-\chi^{\prime}}{\chi\chi^{\prime}}\frac{D_+\left[a(\chi^{\prime})\right]}{a(\chi^{\prime})}j_{\ell}(k\chi^{\prime}),
\end{align}
which describe the redshift distribution $n_z(z)$ of the lensed galaxies, the conditional probability $p(z_{\mathrm{p}}|z)$ for estimating the redshift $z_p$ instead of the true redshift $z$, and the lensing efficiency, respectively.
 
In contrast to the coupling of radial modes due to the lensing efficiency function or to the description of redshift estimation errors, we neglect couplings between the angular wave numbers $\ell$ and $m$ due to incomplete sky coverage, but rather scale our error forecasts with a factor $\sqrt{f_\mathrm{sky}}$. Methods for dealing with survey geometry in the context of the 3dWL-formalism are being developed \citep{leistedt_3d_2015}. The coupling introduces correlations between otherwise uncorrelated modes which reflect the measurement processes and the loss of information due to imperfections that come with it. 

The weak lensing formalism is based on the assumption that the observed ellipticity~$\epsilon$ of a galaxy is the sum of the shear~$\gamma$ and its intrinsic ellipticity~$\epsilon_{\mathrm{s}}$. The intrinsic shapes of source galaxies are therefore a source of noise, given by
\begin{multline}\label{eq:shear_noise}
  \left\langle\bar{\gamma}_{\ell m}(k)\bar{\gamma}_{\ell^{\prime}m^{\prime}}^*(k^{\prime})\right\rangle_{\mathrm{SN}}=\frac{\sigma_{\epsilon}^2}{2\pi^2}\delta_{\ell\ell^{\prime}}^{\mathrm{K}}\delta_{mm^{\prime}}^{\mathrm{K}}\\\times
  \int\mathrm{d}z\,n_z(z)j_{\ell}\left[k\chi^0(z)\right]j_{\ell^{\prime}}\left[k^{\prime}\chi^0(z)\right].
\end{multline}
The intrinsic ellipticity dispersion~$\sigma_{\epsilon}$ is defined by~$\langle\lvert\epsilon_{\mathrm{s}}\rvert^2\rangle=\sigma_{\epsilon}^2$; we assume a value of~$\sigma_{\epsilon}=0.3$ \citep[cf.][]{2003MNRAS.343.1327H}. This expression for the noise holds only if the intrinsic ellipticities of galaxies are uncorrelated, i.e. in absence of intrinsic alignments (see Sec.~\ref{sec:discussion}).

\subsection{iSW signal}
The iSW effect arises as CMB photons propagate through time-evolving gravitational potentials. The difference between the blueshift experienced by a photon entering a gravitational well and the redshift of the exiting photon results in a net frequency change \citep[e.g.][]{1994PhRvD..50..627H}. The sum of these frequency changes along the entire path of the photon causes a temperature shift, which varies with direction according to the evolution of structures along the particular line-of-sight. Since only the integrated frequency change can be observed, the effect contains no distance
information and is analysed only in terms of angular modes. Observations are limited to large scales, as the power spectrum of the gravitational potential is proportional to~$k^{-4}P_{\delta}(k)$ and thus falls off like~$k^{-7}$ on small scales.

The relative temperature change due to the iSW effect is given by the line of sight-integral
\begin{equation}
  \tau(\hat{\mathbf{n}}) = \frac{2}{c^2}\int_0^{\chi_{\mathrm{H}}}\mathrm{d}\chi\,\frac{\partial}{\partial(c\eta)}\Phi\left[\chi(\eta),\hat{\mathbf{n}}\right].
\end{equation}
Here the derivative is taken with respect to conformal time~$\eta$; it can be reparametrized using~$\mathrm{d}\eta=\mathrm{d}t/a=\mathrm{d}a/a^2H(a)$. Transforming from position space to modes~$\Phi_{\ell m}(k;\chi)$ and using Poisson's equation to introduce the density contrast~$\delta_{\ell m}(k;\chi)$, one arrives at the following expression for the amplitude of the iSW effect:
\begin{equation}\label{eq:isw_modes}
  \tau_{\ell m}=-\sqrt{\frac{2}{\pi}}\frac{3\Omega_{\mathrm{m}}}{\chi_{\mathrm{H}}^3}\int_0^{\chi_{\mathrm{H}}}\mathrm{d}\chi\,a^2E(a)\int\mathrm{d}k\,\frac{\partial}{\partial a}\left[\frac{\delta_{\ell m}(k;\chi)}{a}\right]j_{\ell}(k\chi)
\end{equation}
with $E(a)=H(a)/H_0$. In this equation,~$a$ always denotes the scale factor corresponding to the comoving distance~$\chi$ acting as the integration variable. It is obvious that the iSW effect cannot arise during the matter-dominated epoch: In an Einstein-de-Sitter universe with~$\Omega_{\mathrm{m}}=1$, the growth of overdensities in the linear regime is proportional to the scale factor,~$\delta\sim a$, across all scales. For the linear regime the derivative appearing in Eq.~\eqref{eq:isw_modes} simplifies to
\begin{equation}
  \frac{\partial}{\partial a}\frac{\delta_{\ell m}(k;\chi)}{a}=\delta_{\ell m}^0(k)\frac{\partial}{\partial a}\frac{D_+(a)}{a}.
\end{equation}
For the derivative to be nonzero, a significant contribution to the energy density must be due to radiation, curvature, the cosmological constant or dark energy. Given that the Universe appears to be flat and that matter-radiation equality occurs before decoupling, the observed iSW effect must reflect the comparatively recent rise in the dark energy contribution. It is therefore an excellent probe for the equation of state of dark energy
\citep[e.g.][]{2003astro.ph..7335S, 2004ApJ...608...10N, 2006MNRAS.372L..23C, 2006MNRAS.365..891V, 2007MNRAS.376.1211M}. An alternative probe of structure formation at high redshifts is CMB lensing, which collects information about the weak lensing deflection field and the amplitude of structures at redshifts above one, thereby adding information about dark energy \citep{2015MNRAS.449.2205K}. CMB lensing can be measured through its homogeneity-breaking properties by estimating correlations between modes of the temperature or polarisation field at different multipoles. Typical signal strengths are larger than the iSW effect, but the measurement has more complicated noise properties compared to the straightforward cosmic variance inherent to iSW signals \citep{2001ApJ...557L..79H,2002ApJ...574..566H}.

\subsection{Covariance matrix}
In the 3d-approach, the results for the auto-correlation of the iSW effect and its cross-correlation with cosmic shear are as follows:
\begin{equation}\label{eq:isw_auto}
  \left\langle\tau_{\ell m}\tau_{\ell^{\prime}m^{\prime}}^*\right\rangle=\frac{9\Omega_{\mathrm{m}}^2}{2\pi\chi_{\mathrm{H}}^4}\int\mathrm{d}k\,\frac{P_{\delta}^0(k)}{k^2}W_{\ell}^2(k) \, \delta_{\ell\ell^{\prime}}^{\mathrm{K}}\delta_{mm^{\prime}}^{\mathrm{K}},
\end{equation}
\begin{multline}\label{eq:isw_cross}
\left\langle\bar{\gamma}_{\ell m}(k)\tau_{\ell^{\prime}m^{\prime}}^*\right\rangle=\frac{9\Omega_{\mathrm{m}}^2}{8\pi^2\chi_{\mathrm{H}}^4}\sqrt{\frac{2(\ell+2)!}{\pi(\ell-2)!}}\\\times
\int\mathrm{d}\tilde{k}\,\frac{P_{\delta}^0(\tilde{k})}{\tilde{k}^2}G_{\ell}(k,\tilde{k})W_{\ell}(\tilde{k})\, \delta_{\ell\ell^{\prime}}^{\mathrm{K}}\delta_{mm^{\prime}}^{\mathrm{K}},
\end{multline}
with
\begin{equation}\label{eq:covariance_W}
  W_{\ell}(k)=\frac{2}{\chi_{\mathrm{H}}}\int_0^{\chi_{\mathrm{H}}}\mathrm{d}\chi\,E(a)\left[a\frac{\partial}{\partial a}D_+(a)-D_+(a)\right]j_{\ell}(k\chi).
\end{equation}
As the iSW effect manifests itself in changes in the CMB temperature, the noise term is simply given by the CMB spectrum itself,~$C_{\ell}^{\Theta\Theta}$. Since there are no correlations between the temperature fluctuations or the instrumental noise of the CMB observation and the shape noise of weak lensing, no noise term needs to be added to the cross-covariance. Due to these peculiarities, the main limitation for the iSW effect is cosmic variance.

For the full set of iSW and cosmic shear data, the covariance matrix for a given~$\ell$-mode then has the following block structure, with the cross-correlation forming a row- and column-vector indexed by wave number:
\begin{equation}\label{eq:covariance_matrix}
  C_{\ell}(k_i,k_j)=\left(
    \begin{array}{c|ccc}
      \left\langle\lvert\tau_{\ell m}\rvert^2\right\rangle & & \left\langle\tau_{\ell m}\bar{\gamma}_{\ell m}^*(k_j)\right\rangle & \\
      \hline
      & & & \\
      \left\langle\bar{\gamma}_{\ell m}(k_i)\tau_{\ell m}^*\right\rangle & & \left\langle\bar{\gamma}_{\ell m}(k_i)\bar{\gamma}_{\ell m}^*(k_j)\right\rangle & \\
      & & &
    \end{array}
  \right).
\end{equation}
 Ultimately we will combine the weak lensing spectrum including the iSW cross-correlation with a CMB prior: This effectively corresponds to replacing $\left\langle\lvert\tau_{\ell m}\rvert^2\right\rangle$ with the full CMB spectrum, and extending it by inclusion of CMB polarisation spectra. This is true in the limit of no further correlations between the CMB and the weak lensing structure.

\section{Cross-spectra}\label{sec:spectra}

\subsection{Observational scenarios}\label{subsec:obs_scenarios}
We consider two weak lensing surveys. In the framework of the Dark Energy Task Force \citep[][]{2006astro.ph..9591A} these represent research stages III and IV. The specifications for the stage~III experiment are based on the design of the ground-based Dark Energy Survey \citep[DES;][]{2005astro.ph.10346T} with a field size of~$\Omega_{\mathrm{sky}}=5\times10^3\,\mathrm{deg}^2$ and a galaxy density of~$n_0=12\,\mathrm{arcmin}^{-2}$ around a median redshift of~$z_{\mathrm{m}}=0.7$. For the stage~IV survey we choose the parameters~$\Omega_{\mathrm{sky}}=2\times10^4\,\mathrm{deg}^2$,~$n_0=40\,\mathrm{arcmin}^{-2}$ and~$z_{\mathrm{m}}=0.9$ reflecting the goals for the upcoming space-based \textit{Euclid} mission \citep{2011arXiv1110.3193L}. In the following we refer to these survey designs either by their stage or as \lq DES-like/\textit{Euclid}-like\rq.  For the source distribution we follow \citet{1993MNRAS.265..145B} and choose the shape
\begin{equation}
  n_z(z)=6\pi n_0\left(\frac{\sqrt{2}}{z_{\mathrm{m}}}\right)^3z^2\exp\left[-\left(\frac{\sqrt{2}z}{z_{\mathrm{m}}}\right)^{1.5}\right],
\end{equation}
where~$z_{\mathrm{m}}$ is the median redshift of the survey. The distribution is normalised to match the observed redshift-integrated source density~$n_0$. Redshift estimates are assumed to be unbiased with a Gaussian variance,
\begin{equation}
  p(z_{\mathrm{p}}|z)=\frac{1}{\sqrt{2\pi}\sigma(z)}\exp\left[-\frac{(z_{\mathrm{p}}-z)^2}{2\sigma^2(z)}\right]
\end{equation}
with~$\sigma(z)=\sigma_z(1+z)$. We set~$\sigma_z=0.05$ ($0.03$) for the stage~III (IV) survey. Catastrophic outliers \citep[as investigated by][]{2008MNRAS.387..969A} are not contained in the Gaussian ansatz.

Some care must be taken in choosing the maximum wave numbers. The description presented here is only valid for modes which evolve in the linear regime; moreover, \citet{2013MNRAS.434.1808M} showed for the 3d-formalism that the relative contribution of intrinsic alignments (see Sec.~\ref{sec:discussion}) to the covariance increases with the multipole order, in particular for cross-correlations between intrinsic alignments and weak lensing. Realising that the linear approximation breaks down when the density contrast approaches unity, one can estimate the scale or equivalently wave number~$k_{\mathrm{nl}}$ where nonlinear corrections become appreciable. Typically, $k_{\mathrm{nl}}$ is found to be about~$0.3\,h\,\mathrm{Mpc}^{-1}$ today \citep[e.g.][]{1996MNRAS.280L..19P,2003MNRAS.341.1311S}. 
Limber's approximation \citep{1953ApJ...117..134L,1954ApJ...119..655L} then establishes a link to multipole orders~$\ell$: if an observable is projected onto the sphere with a narrow weight function centred on the radius~$R$, the multipole~$\ell$ of the angular spectrum reflects the three-dimensional power at~$k\simeq\ell/R$. Given a survey depth of the order of~$1$--$2\,h^{-1}\,\mathrm{Gpc}$ we restrict the analysis to~$\ell_{\mathrm{max}}=1000$, avoiding the uncertainties of non-linear growth and limiting the numerical expense, and include 500 radial modes up to~$k_{\mathrm{max}}=1\,h\,\mathrm{Mpc}^{-1}$; while the evolution may be mildly non-linear for the smallest radial modes, the main contribution to the highest multipole~$\ell_{\max}$ comes from somewhat lower~$k$-values \citep[cf.][]{2003MNRAS.343.1327H,2005PhRvD..72b3516C}. We discuss the impact of non-linear structure growth in Sec.~\ref{sec:discussion}. The iSW signal is added for multipoles up to~$\ell=40$; from the results in \citet{2014A&A...571A..19P} and \citet{2015arXiv150201595P} it is evident that contributions from smaller scales are negligible.

\subsection{Correlations}
To illustrate the cross-correlations, we calculate Pearson's correlation coefficient
\begin{equation}
  r_{\ell}(k)=\frac{\left\langle\tau_{\ell m}\bar{\gamma}_{\ell m}^*(k)\right\rangle}{\sqrt{\left\langle\lvert\tau_{\ell m}\rvert^2\right\rangle\left\langle\lvert\bar{\gamma}_{\ell m}\rvert^2(k)\right\rangle}}.
\end{equation}
It follows from the Cauchy-Schwarz inequality that~$-1\leq r_{\ell}(k)\leq1$.

\begin{figure}
  \centering
  \includegraphics[width=\columnwidth]{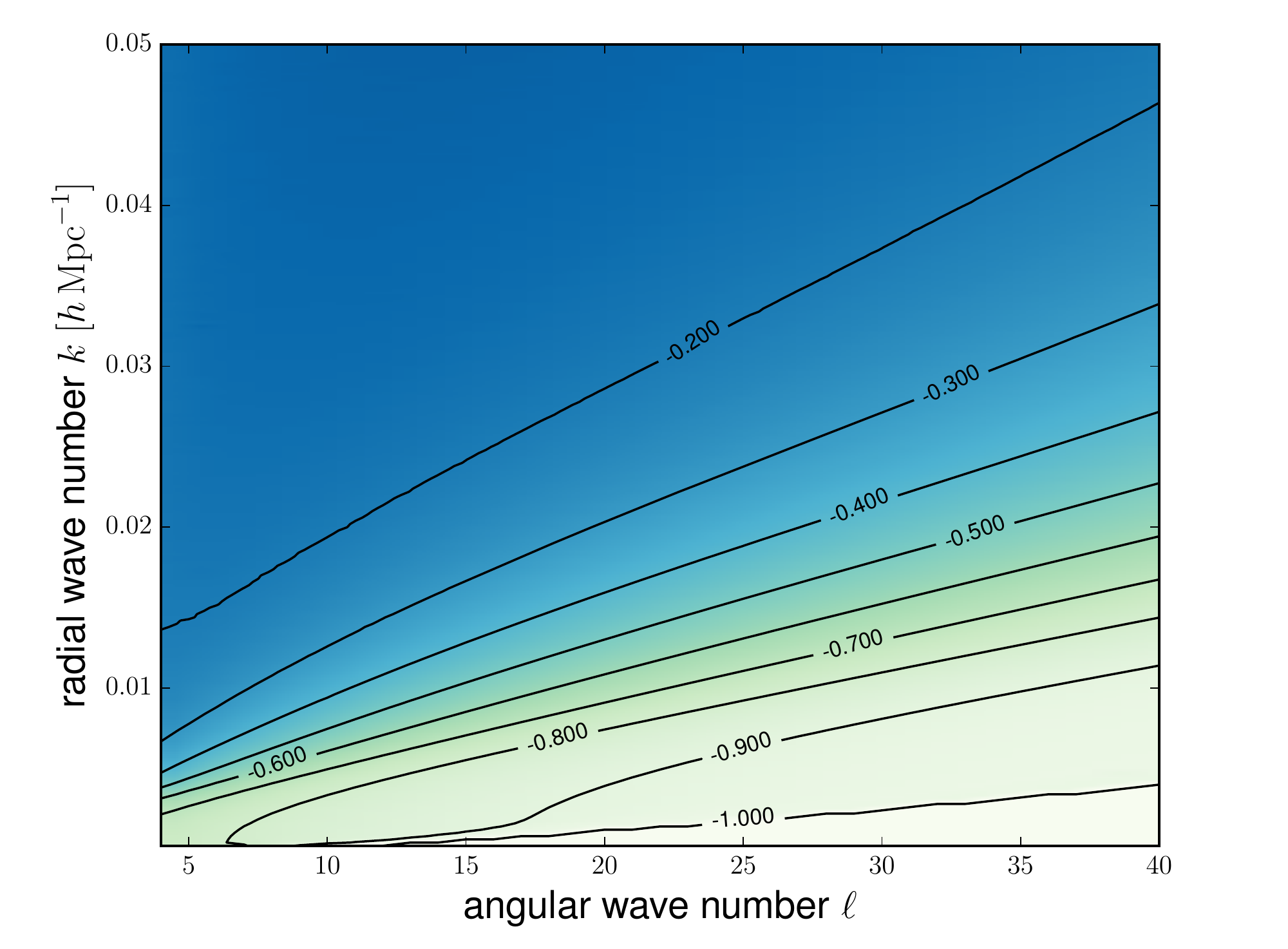}
  \caption{ Pearson coefficient~$r_{\ell}(k)$ for the cross-correlation of the iSW modes and 3d-cosmic shear.}
  \label{fig:pearson}
\end{figure}

Values of the correlation coefficient~$r_{\ell}(k)$ for cosmic shear and the iSW effect are plotted in Fig.~\ref{fig:pearson}. Note that the negative sign mainly reflects the definition of the temperature fluctuation associated with the iSW effect. Finite correlations clearly exist between the two signals: In the limit~$kR\ll\ell$ (where~$R$ again stands for the survey depth)~$r_{\ell}(k)$ is close to~$-1$, implying that corresponding lensing modes contain little independent information compared to the multipole~$\ell$ of the iSW effect. For~$kR\gg\ell$, in contrast, correlations nearly vanish. Cosmic shear modes with large radial wave numbers probe small scales and are therefore only weakly correlated with the iSW signal, which is restricted to the low multipoles. Given this cross-correlation profile, independent information can be gained from the combination of the two data sets, in particular from the low multipoles.

\subsection{Signal-to-noise ratio}
\begin{figure}
  \centering
  \includegraphics[width=\columnwidth]{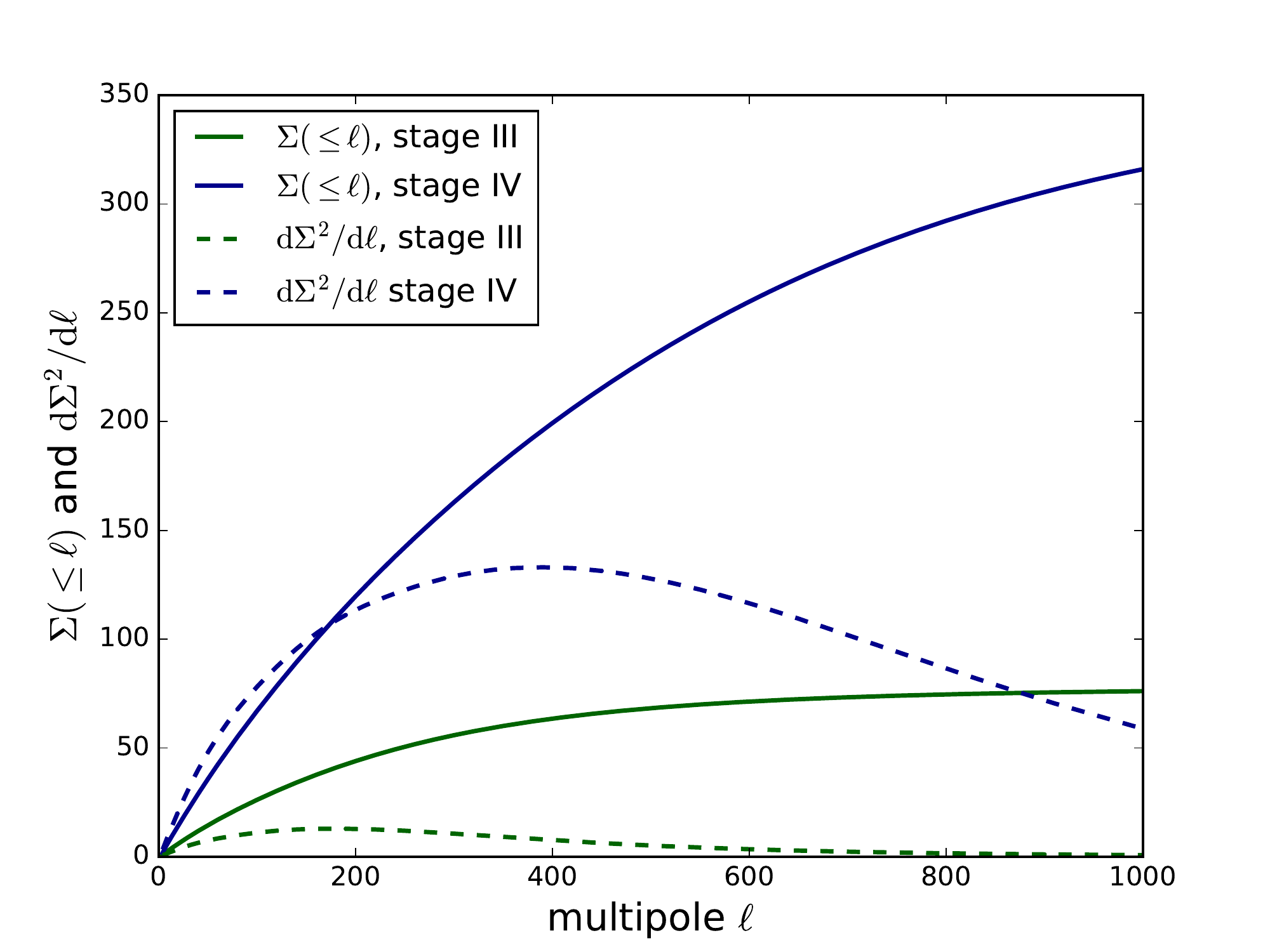}
  \caption{Signal-to-noise ratio as a function of the maximum multipole order~$\ell$ for two weak lensing surveys. Solid lines show the cumulative signal-to-noise ratio~$\Sigma(\leq\ell)$. The dashed lines illustrate the contributions of individual modes.}
  \label{fig:snr}
\end{figure}

The signal-to-noise ratio can be estimated as
\begin{equation}\label{eq:snr}
 \Sigma^2=f_{\mathrm{sky}}\sum_{\ell}\frac{2\ell+1}{2}\operatorname{Tr}\left(C_{\ell}^{-1}S_{\ell}\right)^2.
\end{equation}
The choice of this estimator is motivated by the Fisher matrix approach (see Sec.~\ref{sec:fisher} below).~$S_{\ell}$ denotes the noise-free covariance matrix. The sky coverage~$f_{\mathrm{sky}}=\Omega_{\mathrm{sky}}/(4\pi)$ reflects the reduced information content of a limited field compared to a full-sky survey. We plot the result for both surveys in Fig.~\ref{fig:snr}, using the shorthand notation~$\Sigma(\leq\ell)$ for the cumulative signal-to-noise ratio up to the maximum multipole order~$\ell$, which corresponds to truncating the sum in equation~\eqref{eq:snr} at this value. We also plot the summands to illustrate the contributions of individual~$\ell$-modes.

Comparing the two surveys described in Sec.~\ref{subsec:obs_scenarios} we find that the overall signal-to-noise ratios up to the maximum multipole order~$\ell_{\max}=1000$ differ roughly by a factor of four. This is close to the scaling of Poisson noise with the square root of the number of sources, which is proportional to the field size~$\Omega_{\mathrm{sky}}$ (or sky coverage~$f_{\mathrm{sky}}$) and the projected source density~$n_0$. The exact result reflects additional differences, such as the shape of the source distribution for different depths and the distribution of photometric redshifts. For the realisation of a stage~III survey considered here, the cumulative signal-to-noise ratio flattens off considerably already at~$\ell\approx600$, whereas it is still rising at~$\ell=1000$ for the \textit{Euclid}-like survey. This suggests that an appropriate treatment of higher modes, taking difficulties such as non-linearities and intrinsic alignments into account, could lead to a significant gain in information. The influence of the iSW information on the total significance is below the per cent level; as we demonstrate below, cosmological constraints are nevertheless tightened because derivatives with respect to cosmological parameters are changed and degeneracies broken.

\section{Error forecasts}

\subsection{The Fisher matrix approach}\label{sec:fisher}
The Fisher matrix formalism \citep{1996ApJ...465...34V, 1997ApJ...480...22T} provides error forecasts for parameter estimation problems. Assuming a Gaussian distribution for the data and a Gaussian posterior for the parameters, the Fisher matrix contains lower limits on the variances and covariances of the parameters for a given experimental design or for a model to be examined. The Fisher formalism assumes a Gaussian likelihood $\mathcal{L}$ centered on the reference model and estimates expected statistical errors on parameters and their correlations through the second derivatives of the logarithmic likelihood,
\begin{equation}
  F_{\alpha\beta}=-\left\langle\frac{\partial^2\ln\mathcal{L}}{\partial\theta_{\alpha}\partial\theta_{\beta}}\right\rangle,
\end{equation}
at the maximum of $\mathcal{L}$. A lower bound for the marginal error on the parameter $\theta_{\alpha}$ is then given by the Cram\' er-Rao relation $\Delta\theta_{\alpha}\geq\sqrt{F_{\alpha\alpha}^{-1}}$.

Setting up a Gaussian distribution of measured modes $\tau_{\ell m}$ and $\gamma_{\ell  m}(k)$ with the covariance matrix $C_\ell(k_i,k_j)$, which itself is determined through the cosmological model, yields the Fisher-matrix \citep{1997ApJ...480...22T}
\begin{equation}\label{eq:Fisher_sum}
  F_{\alpha\beta}=f_{\mathrm{sky}}\sum_{\ell}\frac{(2\ell+1)}{2}\operatorname{Tr}\left[C_{\ell}^{-1}C_{\ell,\alpha}C_{\ell}^{-1}C_{\ell,\beta}\right]
\end{equation}
for the case that the total likelihood factorises in terms of the wave numbers $\ell$ and $m$, with a number $2\ell+1$ of statistically independent $m$-modes for each $\ell$. The assumption of a Gaussian likelihood is justified for strong signals such as weak lensing or the spectrum of CMB anisotropies with models of the complexity of $w$CDM, which has been verified through comparisons with MCMC evaluations of the likelihood $\mathcal{L}$ \citep{2012JCAP...09..009W}.  

\subsection{Error forecasts}

\begin{table*}
\begin{minipage}{107mm}
  \caption{Marginal~1-$\sigma$ errors on cosmological parameters from three-dimensional weak lensing ($3d$WL) combined with an iSW experiment. Also shown is the dark energy figure-of-merit. Correlations between data sets are included for~$3d$WL$\times$iSW.}
  \label{tab:errors_isw}
  \begin{tabular}{@{}lrrrrrr@{}}
    \hline
    & \multicolumn{3}{c}{stage~III} & \multicolumn{3}{c}{stage~IV} \\
    Parameter & $3d$WL & $3d$WL+iSW & $3d$WL$\times$iSW & $3d$WL & $3d$WL+iSW & $3d$WL$\times$iSW \\
    \hline
    $\Omega_{\mathrm{m}}$ & 0.172 & 0.149 & 0.104 & 0.0299 & 0.0297 & 0.0273 \\
    $\sigma_8$  & 0.321 & 0.270 & 0.166 & 0.0464 & 0.0460 & 0.0411 \\
    $w_0$  & 1.12 & 1.04 & 0.88 & 0.234 & 0.233 & 0.220 \\
    $w_a$  & 6.07 & 5.27 & 3.70 & 1.028 & 1.021 & 0.927 \\
    $\Omega_{\mathrm{b}}$  & 0.083 & 0.083 & 0.083 & 0.0185 & 0.0185 & 0.0185 \\
    $n_{\mathrm{s}}$  & 0.217 & 0.215 & 0.211 & 0.0458 & 0.0458 & 0.0444 \\
    $h$  & 0.781 & 0.741 & 0.669 & 0.154 & 0.154 & 0.150 \\
    \hline
    FOM & 0.446 & 0.540 & 0.946 & 15.3 & 15.4 & 17.9 \\
    \hline
  \end{tabular}
  \end{minipage}
\end{table*}
We forecast cosmological parameter errors for the following combinations of data sets: \lq$3d$WL\rq{} denotes a cosmic shear survey; the abbreviation \lq$3d$WL+iSW\rq{} stands for independent weak lensing and iSW experiments, for which cross-correlations are ignored and the Fisher matrices are simply added; \lq$3d$WL$\times$iSW\rq{} indicates that the covariance between the two data sets is included. We also generate a CMB prior according to the prescription given by \citet{2006JCAP...10..013P}, considering the auto- and cross-correlations of the temperature and~$E$-modes of the polarisation in a Planck-like experiment, and add it to all of the above combinations. The fiducial cosmology is a $\Lambda$CDM model with the matter density~$\Omega_{\mathrm{m}}=0.314$, fluctuation amplitude~$\sigma_8=0.834$, baryon density~$\Omega_{\mathrm{b}}=0.0486$, spectral index~$n_{\mathrm{s}}=0.962$ and Hubble parameter~$h=0.674$ \citep{2014A&A...571A..16P}. We consider a dark energy component with an equation of state described by the parametrization~$w(a)=w_0+w_a(1-a)$, as suggested by \citet{2001IJMPD..10..213C} and \citet{2003PhRvL..90i1301L}. Matter power spectra are calculated using the \textsc{cosmic linear anisotropy solving system}\footnote{\url{http://class_code.net}} \citep[\textsc{class};][]{2011JCAP...07..034B}.
\begin{figure*}
  \centering
  \includegraphics[width=.9\textwidth]{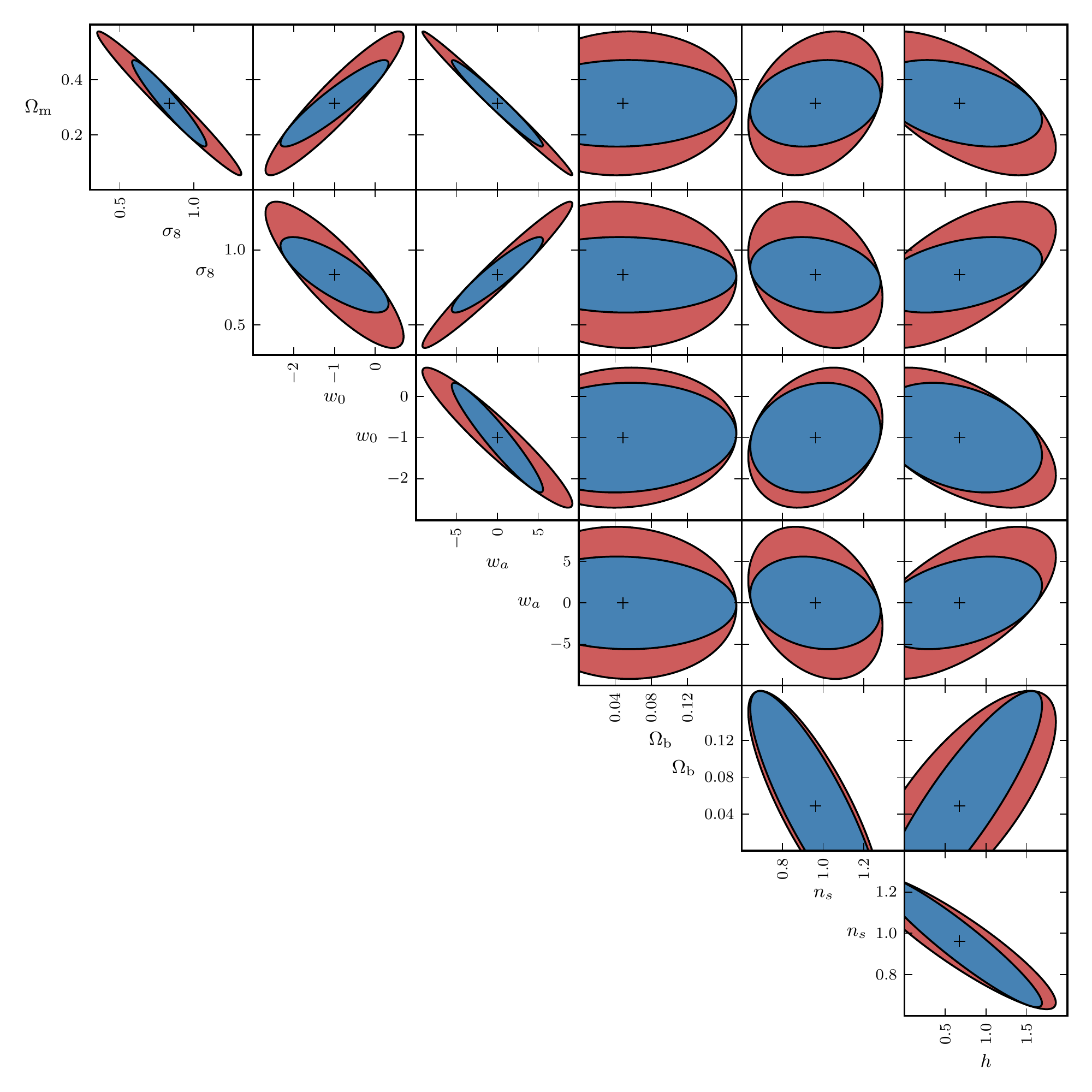}
  \caption{\label{fig:ellipses_des} 1-$\sigma$ confidence regions for cosmological parameters derived from a stage~III weak lensing survey. The red ellipses illustrate constraints from lensing information alone. For the blue ellipses iSW data have been added.}
\end{figure*}
The marginal~1-$\sigma$ errors are presented in Tables~\ref{tab:errors_isw} and~\ref{tab:errors_cmb}. Also included is the figure-of-merit for dark energy (abbreviated \lq FOM\rq), which measures the statistical power of the survey design to test the null hypothesis of a cosmological constant. We adopt the definition by \citet{2009arXiv0901.0721A}:
\begin{equation}\label{eq:fom}
  \mathrm{FOM}=\frac{1}{\operatorname{det}(F^{-1})_{w_0 w_a}}.
\end{equation}
Here~$(F^{-1})_{w_0w_a}$ denotes the~$2\times2$-submatrix of the inverse of the Fisher matrix. The FOM given by equation~\eqref{eq:fom} is inversely proportional to the area of the confidence ellipse in the~$w_0$-$w_a$ plane, so that constraints on the equation of state improve with increasing FOM \citep{amara_figures_2011}.

Table~\ref{tab:errors_isw} shows that constraints are generally poor for the stage~III experiment, implying that it is not possible to constrain such a number of parameters simultaneously for the given survey design. However, it is worth noting that improvements between~$20$ and~$50$ per cent in the marginal errors on~$\Omega_{\mathrm{m}}$,~$\sigma_8$,~$w_0$ and~$w_a$ are achieved when iSW data are added, and the figure-of-merit more than doubles. The influence on~$\Omega_{\mathrm{b}}$,~$n_{\mathrm{s}}$ and~$h$ is much weaker, which is hardly unexpected, as these parameters only affect the shape of the matter power spectrum and degeneracies cannot be broken by additional iSW information. Error ellipses for all parameter pairs, obtained from cosmic shear alone and the full information, respectively, are shown in Fig.~\ref{fig:ellipses_des}.
\begin{figure}
  \centering
  \includegraphics[width=\columnwidth]{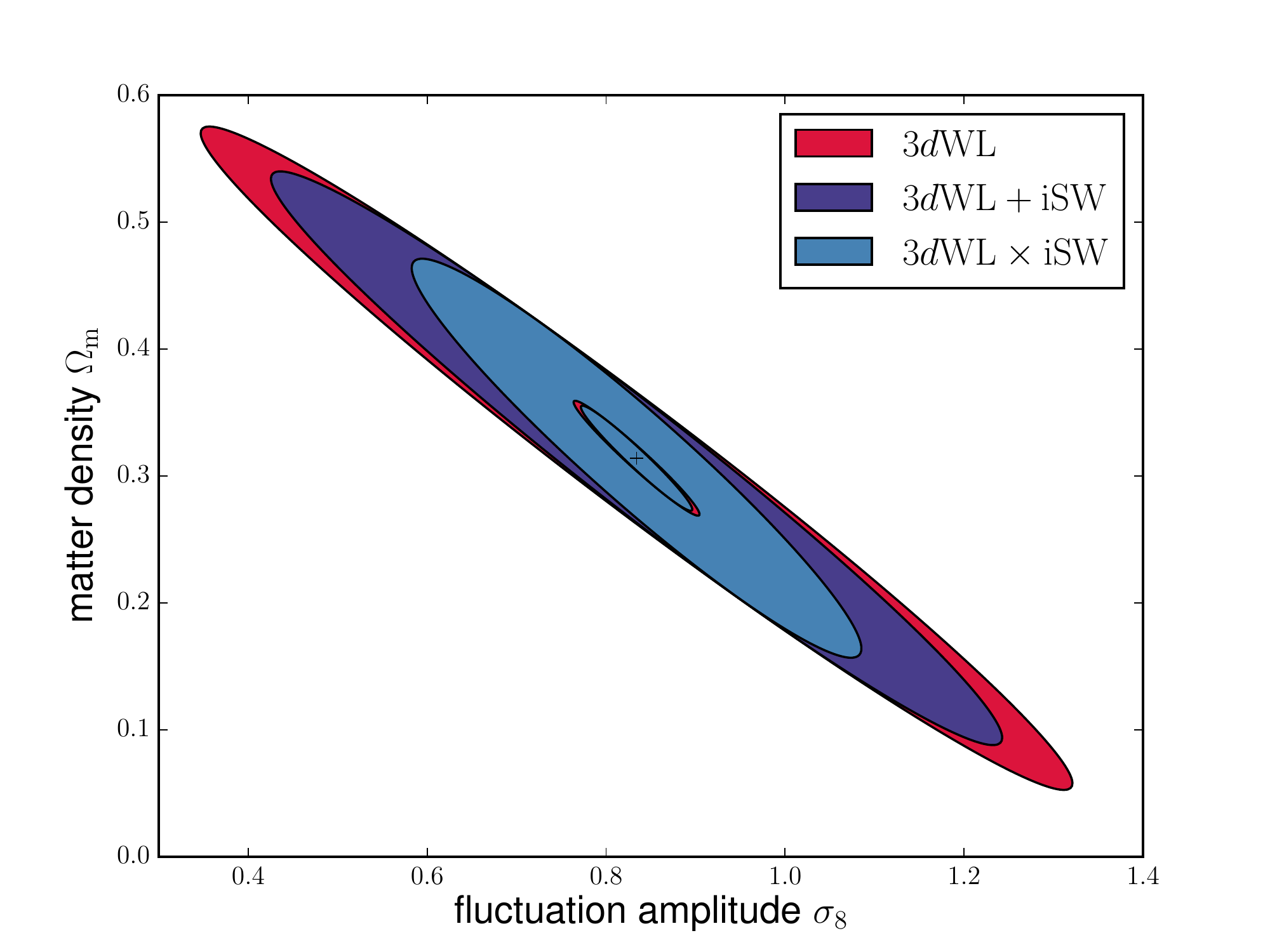}
  \caption{\label{fig:des_Om_w0} 1-$\sigma$ confidence region for the parameter pair~$(\Omega_{\mathrm{m}},\sigma_8)$ from the combination of cosmic shear with a stage~III survey (large ellipses) and the iSW effect, with (\lq$\times$\rq) and without (\lq +\rq) information from cross-correlations, in comparison with a stage~IV survey (small ellipses).}
\end{figure}
\begin{figure}
  \centering
  \includegraphics[width=\columnwidth]{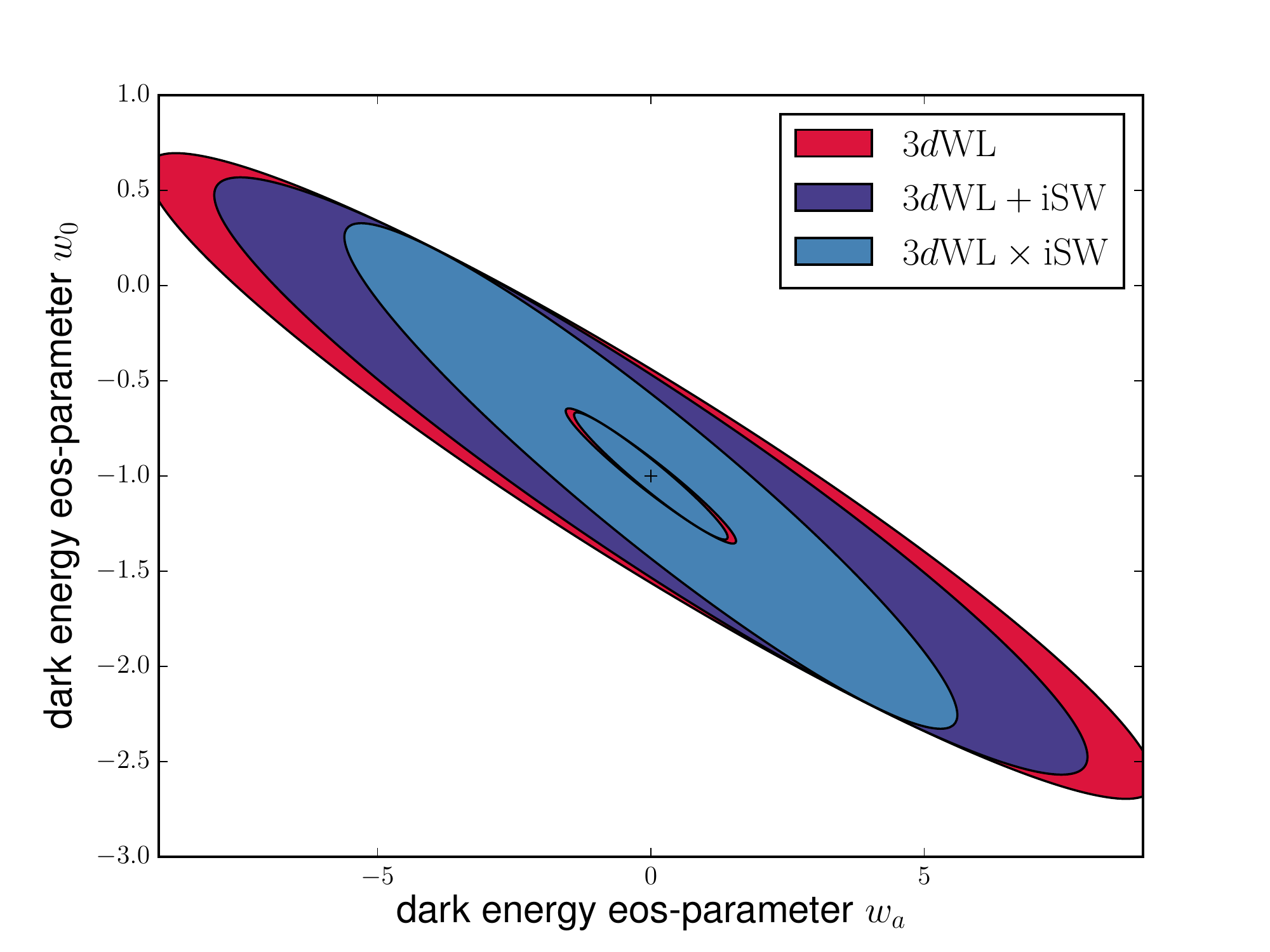}
  \caption{As Fig.~\ref{fig:des_Om_w0}, but for the parameter combination~$(w_0,w_a)$.}
  \label{fig:des_s8_wa}
\end{figure}
If correlations between lensing and iSW data are ignored, errors reduce by five to ten per cent compared to cosmic shear alone; the figure-of-merit increases by~$20$ per cent. This demonstrates that valuable information is encoded in the cross-correlation of the signals, in agreement with our interpretation of the Pearson correlation coefficient above (see Sec.~\ref{sec:spectra}). In Figs.~\ref{fig:des_Om_w0} and~\ref{fig:des_s8_wa} the effect of the additional information on the Fisher ellipses is illustrated, demonstrating the decrease in size and the altered degeneracy.
\begin{figure}
  \centering
  \includegraphics[width=\columnwidth]{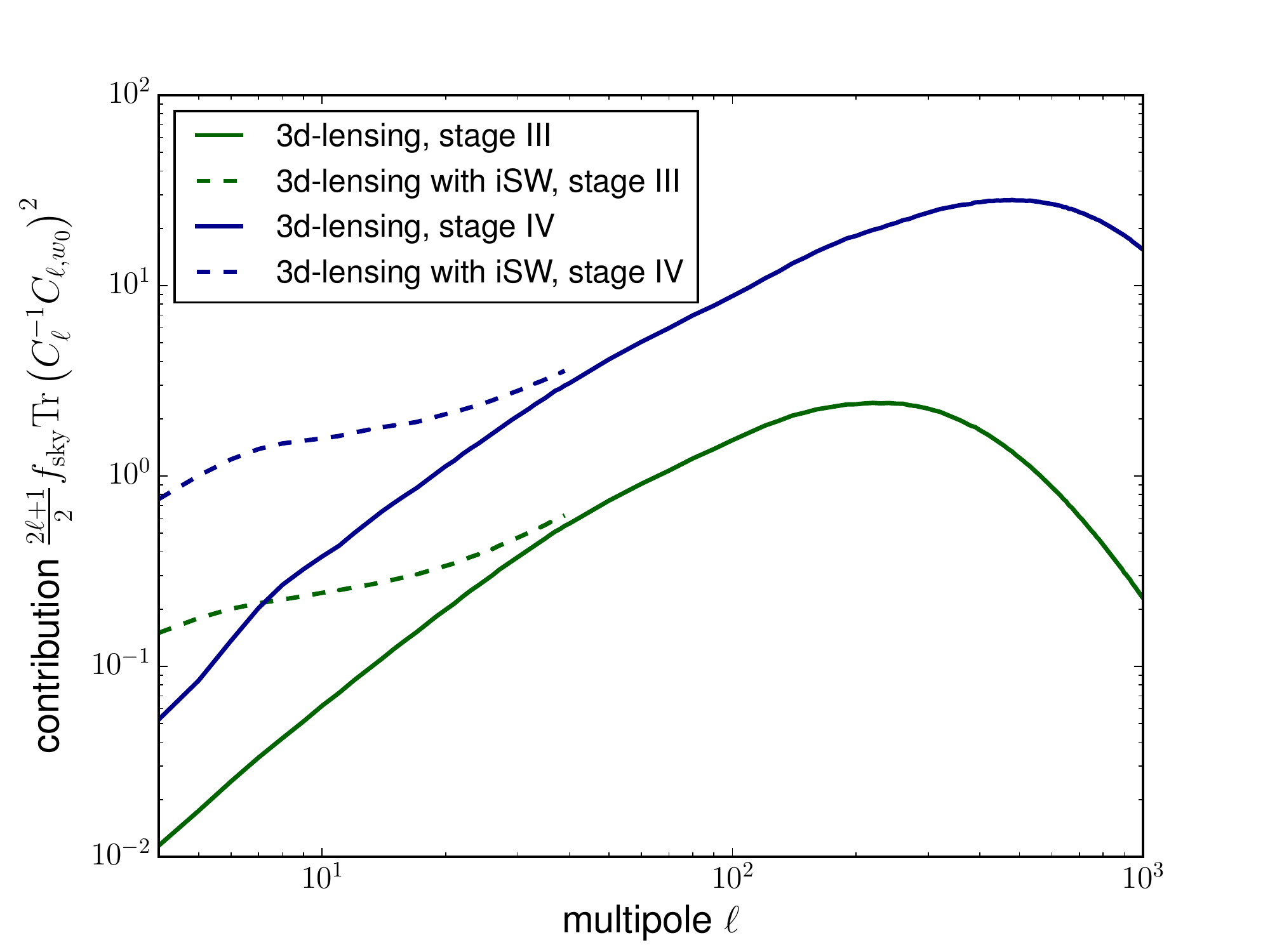}
  \caption{Contributions of different angular scales to the~$w_0w_0$-element of the Fisher matrix for stage~III and stage~IV weak lensing experiments. The curves show the matrix element as a function of the multipole order~$\ell$; the Fisher matrix for the complete information is the sum of these terms over all modes. The solid line is calculated from lensing data only, the dashed line includes iSW measurements.}
  \label{fig:sensitivity_III}
\end{figure}

In Fig.~\ref{fig:sensitivity_III} the contribution of individual angular modes to the full Fisher information is illustrated: we plot the value of the summand in the formula~\eqref{eq:Fisher_sum} for the Fisher matrix, in this example the~$w_0w_0$-element, as a function of the multipole order~$\ell$, which is identical to the Fisher matrix if only a single angular mode is considered. The iSW signal clearly adds power on the largest scales. With growing multipole order the Fisher information in the combined experiment approaches the cosmic shear information; as expected, at~$\ell=40$, the maximum multipole order in the iSW calculation, the gain is almost negligible.

Errors obtained for a \textit{Euclid}-like survey are smaller by about a factor of four (cf. the remarks on the signal-to-noise ratio, Sec.~\ref{sec:spectra}); the figure-of-merit increases by a factor of~$34$. The improvement achieved with iSW information is smaller than for the DES-like survey -- in absence of correlations constraints on~$\Omega_{\mathrm{m}}$,~$\sigma_8$,~$w_0$ and~$w_a$ improve by less than one per cent, whereas errors decrease by about ten per cent if the full covariance is considered; the gain in the figure-of-merit is~$17$ per cent. Figure~\ref{fig:sensitivity_III} suggests that the additional power in the low~$\ell$ modes is overwhelmed somewhat by the cosmological sensitivity of the high lensing multipoles, which dominate the Fisher information. Note that the sensitivity peaks at smaller scales compared to the stage~III survey. It is still high at~$\ell_{\max}=1000$, suggesting that a \textit{Euclid}-like survey can probe even smaller scales; however, an appropriate treatment of non-linear structure formation is indispensable in this case (cf. Sec.~\ref{sec:discussion}).
\begin{table*}
\begin{minipage}{115mm}
  \caption{Marginal~1-$\sigma$ errors on cosmological parameters from a cosmic shear survey and the iSW effect, combined with a CMB prior. Also shown is the dark energy figure-of-merit.}
  \label{tab:errors_cmb}
  \begin{tabular}{@{}lrrrrr@{}}
    \hline
    & & \multicolumn{2}{c}{stage~III} & \multicolumn{2}{c}{stage~IV} \\
    Parameter & CMB & $3d$WL+CMB & $3d$WL$\times$iSW+CMB & $3d$WL+CMB & $3d$WL$\times$iSW+CMB \\
    \hline
    $\Omega_{\mathrm{m}}$ & 0.0139 & 0.0131 & 0.0131 & 0.0082 & 0.0081 \\
    $\sigma_8$  & 0.0133 & 0.0086 & 0.0086 & 0.0060 & 0.0059 \\
    $w_0$ & 0.2356 & 0.2000 & 0.1992 & 0.0937 & 0.0931 \\
    $w_a$ & 0.6588 & 0.5689 & 0.5666 & 0.2412 & 0.2397 \\
    $\Omega_{\mathrm{b}}$ & 0.0021 & 0.0019 & 0.0019 & 0.0012 & 0.0012 \\
    $n_{\mathrm{s}}$ & 0.0027 & 0.0020 & 0.0020 & 0.0018 & 0.0018 \\
    $h$ & 0.0148 & 0.0135 & 0.0135 & 0.0085 & 0.0085 \\
    \hline
    FOM  & 54.1 & 76.3 & 76.6 & 311.6 & 313.9 \\
    \hline
  \end{tabular}
  \end{minipage}
\end{table*}

Table~\ref{tab:errors_cmb} lists errors obtained in combination with a CMB prior. Even a stage~III weak lensing survey is capable of narrowing constraints by~$10$--$30$ per cent compared to the CMB alone; for the stage~IV scenario improvements rise to up to~$60$ per cent. The correlations between cosmic shear and the integrated Sachs-Wolfe effect only affect the errors on~$w_0$ and~$w_a$ below the per cent level. The figure-of-merit increases by~$40$ per cent for the DES-like survey when information from weak lensing and the iSW effect is combined with the prior. For the \textit{Euclid}-like scenario, the gain of a factor of six in the figure-of-merit is remarkable and illustrates the complementarity of CMB and cosmic shear surveys: while marginal errors derived from either of the two data sets alone are of the same order, the different degeneracies between parameters help improve the constraints. Including iSW information, however, does not lead to appreciable further improvements.

\section{Discussion and conclusion}\label{sec:discussion}

We have studied the cross-correlation between weak lensing and the integrated Sachs-Wolfe effect. The iSW effect arises from the evolution of the large-scale structure at late epochs, when the contribution of dark energy (or the cosmological constant) to the density becomes appreciable. This evolution is also probed by cosmic shear, weak lensing of galaxies by the large-scale structure.

In order to exploit the potential of cosmic shear to constrain the growth of structures and spacetime geometry, an analysis of the redshift-dependence of the signal is crucial. Current and upcoming surveys can supply adequate photometric measurements for this purpose. The heightened cosmological sensitivity is the key advantage of a fully three-dimensional formalism over tomographic studies. The numerical complexity, however, increases considerably. In three dimensions, the shear signal can be expanded in a basis of spin-weighted spherical harmonics and spherical Bessel functions. The three-dimensional observable is not statistically homogeneous due to the radial weighting resulting from the distance-dependence of the lensing efficiency, the source-redshift distribution and the dispersion of the photometric redshifts. The loss of statistical homogeneity leads to mode coupling, i.e. finite covariances between radial modes of different wave number.

A challenge for cosmic shear studies is the description of the matter distribution on small scales. In the linear regime, the statistical properties of the density field are preserved, so that correlations between modes can be traced back to the linear matter power spectrum by the appropriate scaling with the growth function. Non-linear growth cannot be represented by such a scaling, since modes no longer evolve independently. Insufficient knowledge of the statistical structure of the matter distribution can bias results for the shear spectrum and cosmological parameters. \citet{1997ApJ...484..560J}, for example, found that second-order statistics of the shear field were strongly affected by non-linearities on angular scales below~$10^{\prime}$, corresponding to multipole orders of~$\ell\gtrsim2000$.  Deviations from a Gaussian likelihood were studied by \citet{2009MNRAS.395.2065T} and \citet{2013PhRvD..87l3538S}. \citet{2006ApJ...640L.119J} showed that the effects of baryons on structure formation, such as radiative cooling and star formation, increased the power spectrum of (two-dimensional) cosmic shear by up to~$10$ per cent for multipoles~$\ell>1000$. \citet{2011MNRAS.415.3649V} similarly found that baryonic feedback lowered the matter power spectrum by a few percent in the range~$0.3\,h\,\mathrm{Mpc}^{-1}\lesssim k\lesssim1\,h\,\mathrm{Mpc}^{-1}$, and \citet{2011MNRAS.417.2020S} interpreted these results with respect to weak lensing tomography, cautioning that parameter biases might be as large as~$40$ per cent.

Intrinsic alignments are an additional uncertainty which becomes increasingly important on small scales \citep[for recent reviews, see][]{2015PhR...558....1T, 2015SSRv..193...67K,2015SSRv..193..139K,2015SSRv..193....1J}. The formalism of cosmic shear is usually built on the assumption that correlations between image ellipticities are exclusively caused by gravitational lensing. Contrary to this hypothesis, finite correlations between the orientations of close pairs of galaxies, known as $II$-alignments, and anti-correlations with the gravitational shear, so-called $GI$-alignments, have been identified in numerical simulations \citep{2000MNRAS.319..649H,2006MNRAS.371..750H,2007ApJ...671.1135K} as well as in observations \citep{2002MNRAS.333..501B,2004MNRAS.347..895H,2007MNRAS.381.1197H,2009ApJ...694..214O}. Both types of alignments are attributed to the influence of the local tidal field on galaxy formation and can mimic a shear signal. The relevance for weak lensing cosmography has been noted e.g. by \citet{2010A&A...523A...1J} and \citet{2010MNRAS.408.1502K}. \citet{2012MNRAS.424.1647K} found the bias in the equation of state parameter~$w_0$ to be as large as~$8$--$20\,\sigma$, but estimates vary depending on the modelling of the spurious contribution to the shear spectrum \citep[cf.][]{2007NJPh....9..444B,2013MNRAS.435..194C}, implying that the impact on cosmological constraints is still far from clear. For three-dimensional cosmic shear, \citet{2013MNRAS.434.1808M} presented auto- and cross-covariances of intrinsic and lensing-induced ellipticities. While~$II$-alignments were shown to be the major contaminant for a \textit{Euclid}-like survey with a median redshift of~$z_{\mathrm{m}}=0.9$, the lensing signal was still an order of magnitude higher at the multipole order $\ell=200$. It is important to emphasise that in contrast to the common argument on statistically uncorrelated noise sources in cross-correlations this would not apply to the iSW effect and intrinsic alignments: As both are sensitive to gravitational potentials and their second derivative, respectively, there is in fact a nonvanishing cross-correlation, which, however, is expected to be small.

We have circumnavigated these obstacles by restricting the analysis to multipoles below $\ell=1000$ and radial modes up to $k=1\,h\,\mathrm{Mpc}^{-1}$. These cuts do not fully but largely exclude the non-linear regime. A more rigorous choice would be $k\simeq0.1\,h\,\mathrm{Mpc}^{-1}$. We do not account for the small deviations that result from applying the linear growth function also in the weakly non-linear regime.
Because of the cuts in harmonic space information from the smallest scales which can be probed by a \textit{Euclid}-like survey becomes inaccessible, but its evaluation would otherwise be subject to substantial uncertainties.

In order to quantitatively explore the impact of the additional information encoded in the iSW signal and its cross-correlation with the three-dimensional cosmic shear field on cosmological parameter errors, we have carried out a Fisher analysis, considering DES-like (research stage~III) and \textit{Euclid}-like (stage~IV) weak lensing surveys. Only the latter can achieve marginal errors of the same order as a CMB experiment, at least on $\Omega_{\mathrm{m}}$, $\sigma_8$, $w_0$ and $w_a$. In light of the aforementioned uncertainties associated with non-linear growth and intrinsic alignments,  including a measurable second order phenomenon in the form of the iSW effect appears more promising than extending the analysis to smaller scales in order to improve constraints. The cosmological sensitivity is clearly raised at the lowest multipole orders. We have demonstrated that the cross-correlation indeed adds valuable information and helps break degeneracies associated with a pure lensing experiment. Marginal errors improve by up to $50$ per cent for stage~III specifications, but they remain large, about a factor of four higher than for the \textit{Euclid}-like survey. For the latter, the inclusion of iSW data tightens constraints on $\Omega_{\mathrm{m}}$, $\sigma_8$, $w_0$ and $w_a$ by about ten per cent and increases the dark energy figure-of-merit by $17$ per cent; the gain is negligible if correlations are ignored. Much stronger constraints can be obtained for both scenarios in combination with other data sets, particularly if a CMB prior is added, largely due to the complementarity of CMB and cosmic shear surveys. Given such a prior and $3d$ weak lensing data, we have shown that the iSW effect in fact does little to lower the errors further. Moreover, CMB data open up the possibility of extending the covariance matrix by the cross-correlation between cosmic shear and CMB lensing, enhancing the cosmological sensitivity even further \citep{2015MNRAS.449.2205K}.

%%%%%%%%%%%%%%%%%%%%%%%%%%%%%%%%%%%%%%%%%%%%%%%%%%



\bibliographystyle{mnras}
\bibliography{references}

% \begin{thebibliography}{99}
% \bibitem[\protect\citeauthoryear{Author}{2012}]{Author2012}
% Author A.~N., 2013, Journal of Improbable Astronomy, 1, 1
% \bibitem[\protect\citeauthoryear{Others}{2013}]{Others2013}
% Others S., 2012, Journal of Interesting Stuff, 17, 198
% \end{thebibliography}

%%%%%%%%%%%%%%%%%%%%%%%%%%%%%%%%%%%%%%%%%%%%%%%%%%

%%%%%%%%%%%%%%%%% APPENDICES %%%%%%%%%%%%%%%%%%%%%

%\appendix

%%%%%%%%%%%%%%%%%%%%%%%%%%%%%%%%%%%%%%%%%%%%%%%%%%

% Don't change these lines
\bsp	% typesetting comment
\label{lastpage}
\end{document}